# Epitaxial growth of graphene-like silicon nano-ribbons

Abdelkader Kara[1,2,*], Christel Léandri[3], Benedicte Ealet[3], Hamid Oughaddou[2], Bernard Aufray[3] and Guy Le Lay[3]

[1]Department of Physics, University of Central Florida, Orlando FL 32816, USA

[2]Département de Physique, Université de Cergy-Pontoise, F-95031 Pontoise cedex, France

[3]CINaM-CNRS, Campus de Luminy, Case 913, F-13288 Marseille Cedex 09, France

*Corresponding author (e-mail : akara@mail.ucf.edu)

Abstract

Graphene, a flat monolayer of carbon atoms tightly packed into a two-dimensional honeycomb lattice (a one atom thick graphite sheet), is prensently the hottest material in nanoscience and nanotechnology. Its challenging hypothetical reflection in the silicon world is coined « *silicene* »; Here, we have demonstrated that the silicon nano-wires self-aligned in a massively parallel array recently observed by STM on Ag(110), are true *silicene* nano-ribbons. Our calculations using density functional theory clearly show that Si atoms tends to form hexagons on top the silver substrate in a honeycomb, graphene-like arrangement.

Pure carbon crystals exist on earth in the form of either cubic diamond gems or simple hexagonal graphite. Graphene, the two-dimensional (2D) one-atom thick single layer building sheet of graphite, is the hottest new material in physics and nanotechnology. It has striking exotic properties, which open potentially novel routes for many applications [1-3]. However, despite the reigning optimism about graphene-based electronics, its discoverers underline that « graphenium » microprocessors are unlikely to appear for the next 20 years [4]. An obvious reason is that replacement of silicon electronics is a tough hurdle. Indeed, silicon-based nanotechnology is highly promising since it is compatible with conventional silicon microtechnology; hence, there has been recently considerable interest in the fabrication, characterization and properties of silicon nanowires (silicon NWs) and nanodots (silicon NDs [5]). *Silicene* [6], the counterpart of graphene in the silicon nano-world has not been found until now, although it has attracted strong theoretical attention for several years [7]. On the experimental side, several papers have reported on the fabrication of silicon nanotubes (silicon NTs) in the laboratory, including most presumably single-walled ones [8,9].

In the following, we give evidence of the controlled epitaxial growth of *silicene* nano-ribbons (*silicene* NRs), which could be a new paradigmatic figure in the silicon realm.

This discovery stems from the search for the intimate atomic structure of the novel one-dimensional (1D) silicon nanostructures (see Fig. 1), found recently in Marseille upon carefully controlled growth of silicon on silver (110) surfaces under ultra-high vacuum conditions [10]. They were observed with scanning tunneling microscopy (STM), and were called previously "silicon nanowires" in the absence of precise structural determination; as we will see, they are one atom thick graphene-like silicon stripes, i.e., true *silicene* NRs. Despite the lack of knowledge on their atomic arrangement, we could study in details their physical and chemical properties, which are exceptionally striking.

First, we mention the straightness and massive parallel alignment along the [-110] channels of the bare surface, giving highly perfect identical (but chiral) quantum-objects in the macroscopic ensemble (just differing in their lengths) over the whole substrate surface (~ 40 mm$^2$). Second, we underline several structural features, namely: (i) the exact "by two" periodicidy along their sides (2 $a_{Ag[-110]}$ nearest neigbourg distances, that is, 0.578 nm;); (ii) the same common height of ~ 0.2 nm; (iii) a « magic » width of ~1.7 nm (i. e., about 4$a_{Ag[001]}$); (iv) a unique electronic signature reflected in the STM images showing a key structure formed by a square and a parallelogram side-by-side; (v) the symmetry breaking across the ribbon with a dip on one side, while the two resulting chiral species phase separate spontaneously into large left-handed and right-handed magnetic-like domains.

Third, we stress the strong metallic character and unprecetented spectral signatures, i.e., quantized states in the sp region of the valence band below the Fermi level with a strict 1D dispersion along the length, and finally, the narrowest Si 2p core-level lines ever reported to our knowledge in the solid state [10,11]. Fourth, we emphasize the surprising burning match mechanism of the oxidation process of these NWs, which starts at their extremities with appearance of several oxidation states similar to those observed upon the formation of the $SiO_2$/Si(111) interface [12]. Finally, we stress that at slightly higher temperatures these unique 1D silicon NWs self-assemble by lateral compaction to form a grating with a pitch of just ~ 2 nm covering the whole substrate surface [13].

Soon after the publication of our first findings [10], a possible model of these silicon NWs was proposed by Guo-min He on the basis of a density-functional theory study; it remains the sole tentative approach to elucidate their elusive atomic structure [14]. The proposed optimum geometry presented interesting features, typically silicon dimers, the building blocks of the Si(100)2x1 surface reconstruction. However, being constructed on a constrained, improper 2x5 rectangular motif, the stability of this geometrical model could be eventually questionned on a larger cell (in fact the STM image in Fig. 1 shows a x4 periodicity along the nanowires). Furthermore, as mentionned above, the oxidation states we have identified since then [12], resemble those found upon formation of the $SiO_2$/Si(111) interface, but not those found upon formation of the $SiO_2$/Si(100) interface, as would have been expected [15]. For such reasons it appeared mandatory to us to re-analyse theoretically the silicon nanowires atomic structure starting not only from energetic criteria but also from our most recent experimental observations and measurements. In the present paper we focus on structural aspects leaving a detailed analysis of the electronic properties to a forthcoming paper.

The three dimensional STM image presented in Fig. 1 shows that the NWs are composed of three rows of protrusions, each with a perfect x2 periodicity, oriented along the [-110] direction of the surface. The one on the right side is shifted from the others by one $a_{Ag[-110]}$ silver parameter, hence, the true overall periodicity is actually x4. This glide gives a transverse asymmetric structure schematically described from left to right as a square (side: ~0.6 nm) joined to a parallelogram, highlighted with black lines on Fig. 1.

This overall geometrical description is the starting point for our theoretical study. Our *ab initio* density functional theory (DFT) [16,17] calculations are based on an approach rather different from that of Guo-min He [14]. A comprehensive study of energetics and electronic structure was made by solving Kohn-Sham equations in a plane-wave basis set using the Vienna *ab initio* simulation package [18-20] (VASP). Exchange-correlation interactions are included within the generalized gradient approximation (GGA) in the Perdew-Burker-Ernzerhof form [21]. The electron-ion interaction is described by the projector augmented wave method in its implementation of Kresse and Joubert [22,23]. A plane-wave energy cut-off of 250 eV was used for all calculations and is found to be sufficient for these systems. The bulk lattice constant for Ag was found to be 0.4175 nm using a k-point mesh of 10×10×10. The slab supercell approach with periodic boundaries is employed to model the surface and the Brillouin zone sampling is based on the technique devised by Monkhorst and Pack [24] with a (3x2x1) mesh. The slab consists of 5 layers of Ag(110) each containing 24 atoms (4x6), a substancial enlargement with respect to the previous theoretical work [14]. The choice of 5 layers was made on the assumption that adsorbtion of the silicon NWs could introduce substantial structural perturbations to the substrate. The whole system is allowed to relax to the optimum configuration (with forces on every atoms less than 0.01 eV/Å), with only the bottom layer atoms kept fixed.

A rigorous determination of the lowest energy configuration of silicon NWs on Ag(110) requires the calculation of all possible arrangements of Si atoms on Ag(110) contained in the surface unit cell. There are two problems with this approach, (i) STM images are not resolved enough to reveal unambiguously the number of Si atoms per unit cell; (ii) even if this number could be known, it would be large (30-40 atoms) and a systematic (DFT based) calculation of the total energy for all possible combinations is computationally very demanding, even impossible in the current situation. We adopt a rather different approach that, indeed, does not guarantee the lowest energy structure, but reveals possible configurations that comply with a maximum set of experimental observations. We recall, here, the set of observations retained for the elimination process; they are: (i) the width of each nano-wire covering about 4 times the lattice constant of Ag; (ii) a periodicity along the edges of the ribbons of twice the nearest number distance; and (iii), a signature in the STM images of the presence of a square attached to a parallelogram.

There are several configurations incorporating the first two observations and we will pay more attention to the third one to perform the final selection. Silicon NWs with a variety of geometries (i.e. number of Si atoms between 27 and 36 per unit cell) were studied. The silicon

NWs starting configurations can be divided in two categories: those, in which Si atoms form hexagons, in a honeycomb arrangement, and those forming rectangles. Let's recall that previously, Guo-min He proposed an arrangement where dimers of Si atoms formed a rectangular lattice on a 2x5 Ag(110) substrate [14]. We have performed calculations for the same arrangement on a 4x6 Ag(110) substrate and found that all the rectangular type geometries experience large deformations upon relaxation, with the Si atoms approaching a hexagonal configuration. The other category of starting configurations consists of aligning hexagons along the direction 30º from the channels of Ag(110) surface. Typically, a Si30 configuration would consist of 4 hexagons while a Si36 would consist of 5 hexagons. As a criterion of selection/elimination, we have used the presence/absence of the prominent signature mentioned above (a square adjacent to a parallelogram) and found out that only the Si30 entity provides clearly, such a signature as seen in Fig. 2(b), where we show the calculated STM image.

In Fig. 3(a), we display the resulting Si30 optimized structure in correspondence with an experimental 3D STM image (Fig. 3(b)) which reveals, upon perusal, that the x4 periodicity of the NRs extends on the bare silver surface itself. We believe that this propagation is the signature of the cross-talk between the nanowires, which is most probably at the origin of the formation of large magnetic-like domains, consisting separately of left-handed and right-handed NWs. Regardless of the starting type of configuration, our calculations show that Si atoms tend to form hexagons on top of the silver substrate in a honeycomb, graphene-like arrangement, in other words the silicon NWs are *true silicene nano-ribbons (silicene NRs)*. This optimized structure of *silicene* ribbons on top of the silver substrate (which presents itself a rectangular structure) together with the nature of the bonding between the different atoms, conspire to provide unique features in the electronic density of the system. This translates the presence of 9 Si atoms out of 30 at the top most part of the *silicene* NRs (nearly at the same height), colored differently in Fig. 4(c). This demonstrates that a system of atoms like these silicon NWs, having local hexagonal symmetry, may indeed conspire in such a way to give the impression of a local square symmetry.

In more detail, after full atomic relaxation, the atomic positions of the silicon atoms are very close to those of a slightly curved *silicene* nano-ribbon: the ribbon creates an asymmetric corrugation in the charge density profile (Fig. 4(b)), in nice accord with the STM profiles (see Fig. 1).

This curvature may favour their stabilization with respect to a purely $sp^2$ bonded flat ribbon through intermediate $sp^2$-$sp^3$ hybridizations. This "arch-shaped" configuration consists of edge Si atoms penetrating the Ag(110) channels for a better "grip" onto the surface and hence creating a "bump" in the STM images (reflecting a protrusion in the charge density). Indeed, at the edges of the *silicene* NRs, there are dips that we will assimilate to "gutters"; along which the energy profile is weakly corrugated. Consequently, we can assume that when Si atoms come directly from the sublimation source or diffuse to the ribbon edge, they are very rapidly dispatched to the ends of the nano-ribbons, contributing to their elongation. We believe that it is this "gutter effect" which is responsible for the formation of very long nano-ribbons at room temperature and above.

The calculated width (from the center of the far left Si atoms to the center of the far right Si ones) is ~1.51 nm, while the STM images show an apparent width of about 1.7 nm (see the profile in Fig. 1). In fact, this is in fairly good agreement if we take into account the van der Waals radii of the Si atoms.

The substrate itself, experiencing buckling, is somewhat altered by the *silicene* NRs. The chain of silver atoms on the left of the nano-ribbon is not straight and shows substancial displacements (about 0.08 nm) from the ideal positions; a new component disclosed in recent high-resolution synchrotron radiation Ag 3d core-level spectroscopy measurements stems from these displaced silver atoms. Note that it is only one side of the ribbon that is hence affected, reflecting again the asymmetry nature of the system, in good accord with the experimental observation. The structural effects of the *silicene* NRs on the silver substrate go as deep as the fourth layer. The bucklings in the $2^d$, $3^d$, and $4^{th}$ layer amount to about 0.06, 0.04 and 0.02 nm, respectively.

In conclusion, we have demonstrated the experimental synthesis of *silicene*, in the form of massively parallel *silicene* nano-ribbons. The arrangement of 30 silicon atoms in a 4x5 unit cell on the Ag(110) surface relaxes into an arch-shaped graphene-like configuration that is consistent with the main experimental observations. Sharing the same atomic geometry with graphene , *silicene* posseses the same striking electronic features: typically, its charge carriers are massless relativistic Dirac fermions [6]. The synthesis of massively parallel one-dimensional *silicene* ribbons offer the possibility to study new intriguing physics, especially through the delicate interplay between spin-charge separation in Luttinger liquid in 1D and Dirac fermions in 2D.

Concerning potential applications, since to make up a circuit with graphene sheets one envisions to slice them wider or narrower and in different patterns depending on wether a wire, a ribbon or some other component are needed, the spontaneous formation of *silicene* NRs, which can further develop into wider, longer and slightly thicker, still massively parallel, *silicene* NWs is rather promising, not to mention the formation of a dense array on the whole substrate surface [10,13]. In this respect, this dense array of massively parallel *silicene* nano-ribbons (giving a 5x4 diffraction pattern) corresponds to an architecture of ~ 5 $10^6$ channels/cm, comparing favorably with the active massively parallel architecture of silver adsorbed silicon lines (in a sense the counterpart system) on silicon carbide surfaces [25].

With reference to graphene, the epitaxial growth process we have initiated is not so far from the method developed by C. Berger *et al*. [26] for mass-production for industrial applications. Very encouraging for practical purposes is the high stability towards oxidation of the SiNTs reported by De Crescenzi *et al*. [8], whose data indicate that the tubes are not significantly oxidized after exposure to air. This behavior probably explains why the oxidation process of the *silicene* NRs starts at their extremities, which are their main defects [12], but not their sides. Ways to tailor the sizes of this *silicene* NRs, extract them and encapsulate them have been envisaged. Typically, in this context, individual *silicene* NRs or NR arrays could be prepared in large scale on thin silver (110) epitaxial films grown, e.g., on GaAs(100) substrates [27]. This opens the most promising route towards wide-ranging applications by further deposition of an insulating support on top and chemical removal of the primary metallic substrate.

**Acknowledgements**

The work of AK was partially supported by a start-up fund from the University of Central Florida. AK thanks the University of Cergy Pontoise and CINaM-CNRS Marseille for support.

FIG. 1. STM image of straight, parallel one dimensional silicon nanostructures grown on a silver (110) surface (6.22 x 6.22 nm$^2$, filled states, 3D view). Line profiles indicate the 2 $a_{Ag[-110]}$ periodicity along each line of protusions ([-110] direction) and the asymmetrical shape along the orthogonal direction. The geometrical aspect of the silicon NWs (a square joined to a parallelogram) is drawn in black.

FIG. 2. Experimental and calculated STM images. (a) STM image of an individual silicon NW showing the structural signature made of a square adjacent to a parallelogram; (b) calculated STM image for the Si 30 configuration.

FIG. 3. *Silicene* nano-ribbons. (a) STM image; (b) corresponding ball model.

FIG. 4. Calculated atomic and electronic structures of *silicene* nano-ribbons. (a) Top view (dark blue: first layer Ag atoms, light blue: second layer Ag atoms; top most Si atoms are in red, other Si atoms forming hexagons are in green); (b) cross section; (c) charge density in cross section.

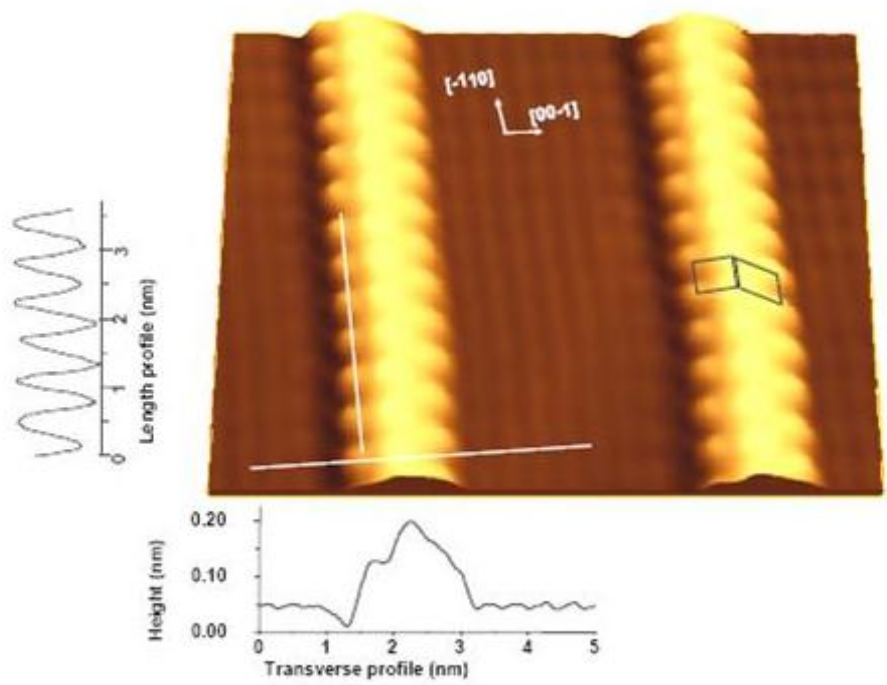

FIG. 1. A. Kara *et al.*

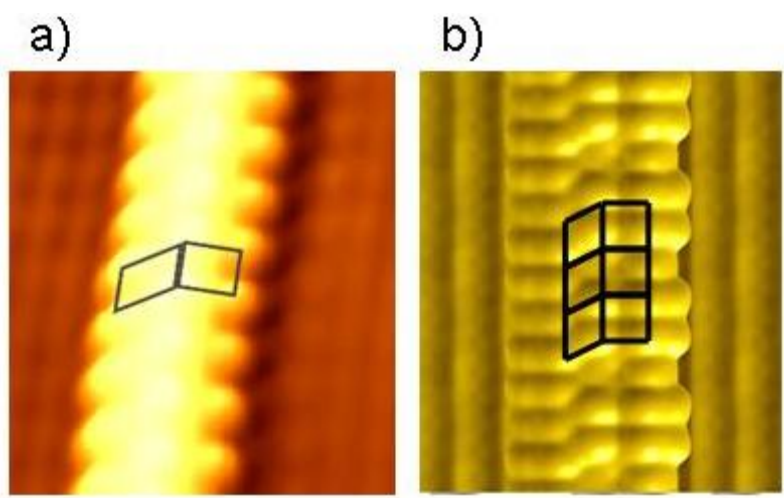

FIG. 2. A. Kara *et al.*

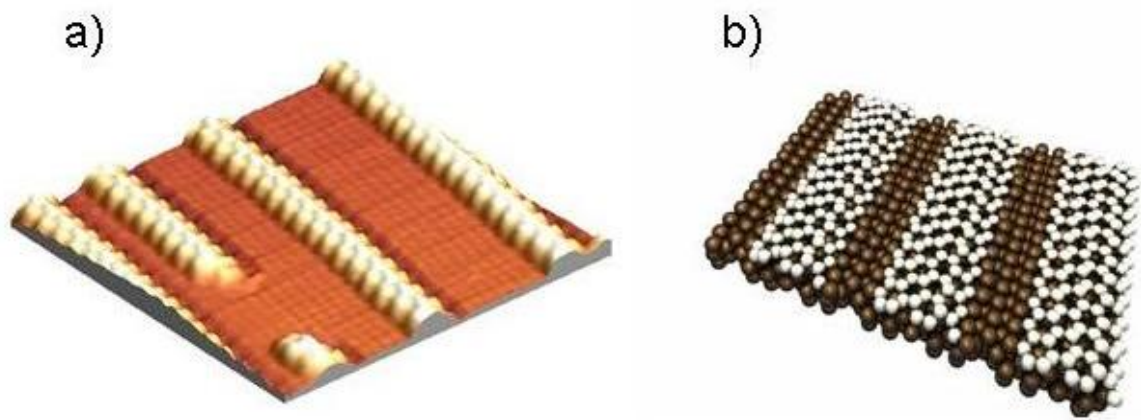

FIG. 3. A. Kara *et al.*

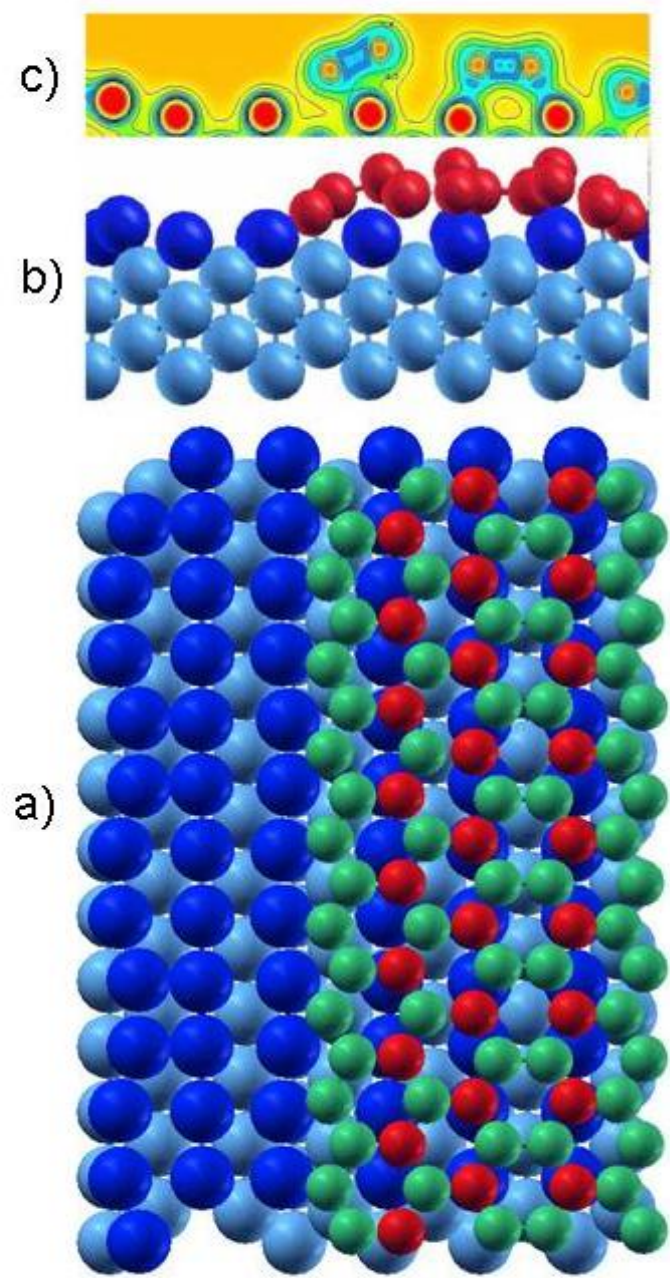

FIG. 4. A. Kara *et al.*